# American society keeps a lid on the number of deaths from guns and car accidents but not from mass shootings

**Theodore Modis**[*]

**ABSTRACT**

The number of deaths from car accidents and from the unlawful use of guns can be described by logistic growth curves. The annual rates of both have traced completed logistic trajectories following which they have been self-regulated for many decades at what seems to be a homeostatic equilibrium level through legislative actions. Exception constitutes the number of deaths from mass shootings, which has been so far tracing an exponential trajectory. Despite the fact that mass-shooting deaths represent only 0.1 percent of all gun deaths today, they are poised to continue growing exponentially until they become the major cause of gun deaths, short of unprecedented action by society.

______________________________
* Growth Dynamics
  Via Selva 8
  6900 Massagno
  Lugano, Switzerland
  E-mail: tmodis@gmail.com

1. Introduction

It was in the late 19th century that English philosopher and sociologist Herbert Spencer suggested that society can be seen as a living organism striving to maintain a balance between its various parts to ensure stability and health.[1] In mid-20th century John Williams, an analyst in Rand Corporation, pointed out that society had been self-regulating car deaths at a flat annual rate for decades despite significant increases in the number and the performance of cars on the roads. He had put it rather crudely: "I am sure that there is, in effect, a desirable level of automobile accidents – desirable, that is, from a broad point of view, in the sense that it is a necessary concomitant of things of greater value to society."[2] Despite the cost in human life, society seems willing to pay a price for the convenience and advantages of using cars.

In this article we will examine also gun deaths from the same perspective. Nicholas Kristof (outspoken journalist on gun control) argues that "We have a model for regulating guns: automobiles" and continues "We don't ban cars, but we work hard to regulate them … so as to reduce the death toll they cause."[3] Guns in the US have things in common with cars. They both serve a useful purpose, they both cause violent deaths of often innocent people, and they both are generally cherished by their owners. Whereas the usefulness from cars may exceed considerably that from guns, it is not a priori obvious what Americans would answer if asked: "Would you rather give up your car or your gun?"

2. Logistic growth of the number of deaths

The effectiveness of efforts to limit the number of deaths resulting from car accidents or from the unlawful use of guns is best reflected in the evolution of the number of annual deaths *per capita* rather than per mile, or per hour of driving, or per car/gun. After all, it is society that feels the pain and tries to do something about it. The overall evolution of the number of deaths follows a trend that can be generally described by an S-shaped natural-growth curve, a *logistic* function:

$$f(t) = \frac{M}{(1+e^{-\alpha(t-t_0)})} \qquad (1)$$

where $M$ the final maximum, $\alpha$ a constant reflecting the slope, and $t_0$ the center of the growth process.

Such curves are used to describe the evolution of species populations growing into ecological niches under conditions of completion (survival of the fittest), e.g., rabbits in a fenced-off grass field. The curve begins exponentially but then slows down as it approaches a ceiling, which reflects the capacity of the niche to accommodate the final size of the population. But the application of such curves has been extended to variables describing the growth of inanimate populations as long as the niche is of finite size and necessitates competition. Many growth processes in society have been shown to follow this pattern including products, markets, primary energies, technologies, personal achievement, accidents, learning, creativity/productivity, and criminality.[4][5]

2.1 Car accidents

Car accidents are annually responsible for approximately 1.3 million deaths worldwide, according to the World Health Organization, and up to ten times as many may suffer injuries. Cars have been likened to murder weapons. From the beginning of car's history auto manufacturers and legislators have been trying to improve car safety and limit the number of accidents.

The graph at the top of Figure 1 shows data for the US from the beginning of the 20[th] century. The number of deaths caused by car accidents per 100,000 inhabitants grew steadily until the mid 1920s when this number reached around 22.5 (thin black line.) This growth reflected the growing number of cars being used by the general population until 1930. But then it stabilized even though the number of cars on the roads and their use kept increasing. From the mid 1920s onward, the number of deaths fluctuated above and below this level, representing some kind of a homeostatic equilibrium. The smooth S-shaped light gray line in Figure 1 is a logistic curve fitted on the data. It begins exponentially but eventually slows down to reach a ceiling of 22.5, which reflects a homeostatic equilibrium with the data oscillating above and below this level thereafter. Table 1 gives the parameters of the logistic as determined by the fit.

Society became sensitive and reacted to deviations from this final level with varied state and federal legislation. Upward excursions may have triggered more safety legislation, or simply sufficient public outcry, that influenced societal attitudes toward safety, while downward excursions may have resulted in the relaxation of speed limits. For example, in 1965 Ralf Nader published a bestseller *Unsafe at Any Speed*, which criticized the automotive industry for its safety record. The book helped lead to the passage of the National Traffic and Motor Vehicle Safety Act in 1966. The peak in car-accident deaths in the late 1960s was largely eliminated by the requirement by the National Highway Safety Bureau (NHSB) and later the National Highway Traffic Safety Administration (NHTSA) to institute safety standards such as equip all new passenger vehicles with seat belts starting in 1968.

What is remarkable is that for more than half a century there has been "self-regulation" of the number of deaths despite significant increases in the number and the performance of cars on the roads, improvements in the technology and legislation around safety, and enhanced public awareness about safe driving.

During recent decades the fatalities from car accidents were significantly reduced, mostly due to intercity travel switching from cars to airplanes. By early 21[st] century this number had reduced to around 10, similar to that from the use of firearms, see dotted line in Figure 1.     The picture worldwide is qualitatively rather similar, with the rate of road traffic deaths remaining fairly constant around 18 per 100,000 polulation between 2000 and 2016.[6]

**2.2  Gun deaths**

The lower part of the graph in Figure 1 (thick black line) shows the situation regarding deaths from the unlawful use of firearms, i.e., homicides plus suicides plus accidents whenever separately reported. If we now apply the same reasoning as with deaths from car accidents, we find that the process can again be described with a logistic curve – smooth dark gray line in Figure 1– which grew to reach a ceiling, a homeostatic equilibrium, at around 10 deaths per 100,000 inhabitants in the early 1930s. Table 1 shows the parameters of the logistic as determined by the fit.

It was the National Firearms Act of 1934 (NFA) which brought this number back down to around 10 following an upward deviation that included excessive criminality such as the gangland crime of the Prohibition era (e.g., the St. Valentine's Day Massacre of 1929) and the attempted assassination of President-elect Franklin D. Roosevelt in 1933. From the early 1930s onward the number fluctuated above and below this level sometimes triggering restrictive legislative actions or at other times loosening gun-control laws. Apparently, the price society was willing to pay for the use of guns (10 deaths per 100,000) was smaller than that for the use of cars (22.5 deaths per 100,000), reflecting perhaps the respective usefulness as perceived at the time. An additional explanation may be found in that gun ownership/usage is generally lower than that of cars.

The drop in gun deaths in the early 1980s cannot be attributed to a single event. It was probably related to varied legislation/regulations across the country on state level.

In more recent times, the Brady Handgun Violence Prevention Act of 1993 (BHVPA) and the Federal Assault Weapons Ban of 1994 (FAWB) were triggered by – and eliminated – the upward deviation of this number around the mid 1990s. But the latter expired on September 13, 2004, thus preparing the ground for a new upward deviation.

What is remarkable is that for an entire century there have been obvious and less obvious mechanisms that regulated this number (the prolonged downward fluctuation around the 1940s was largely war-related.) It seems that around 10 deaths annually per 100,000 is the price society is willing to pay for the use of guns, a necessary concomitant of things perceived by its citizens as of greater value.

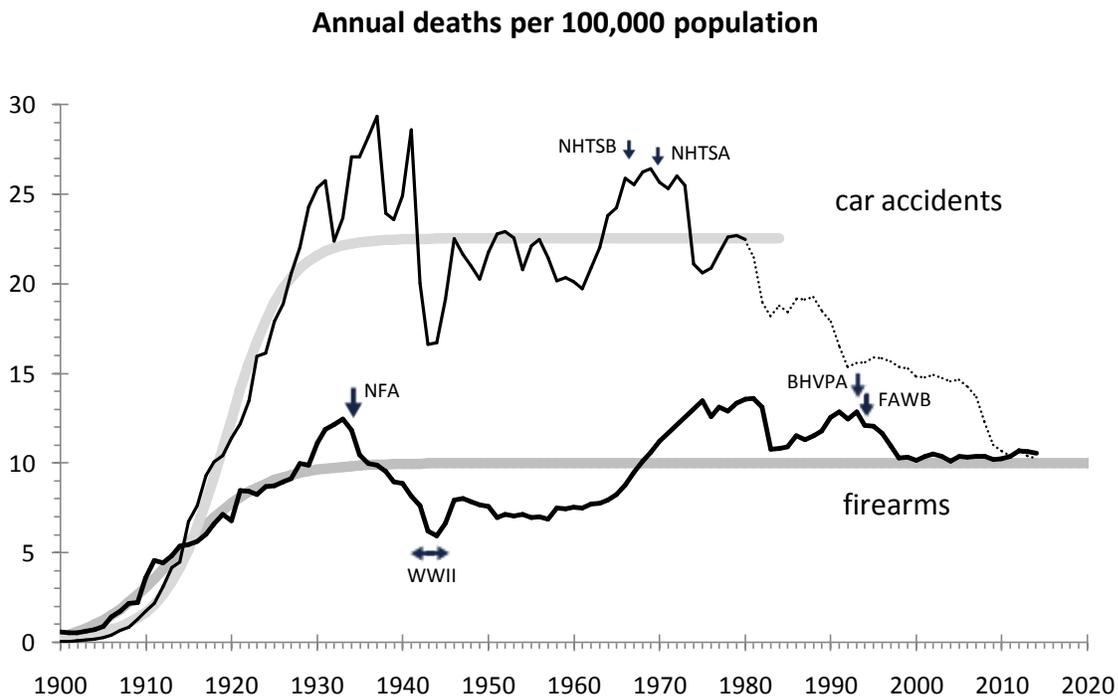

**Annual deaths per 100,000 population**

Data source: Statistical Abstract of the United States[7]

Figure 1. Annual rates of deaths from car accidents and the unlawful use of guns. The smooth gray lines are logistic fits to the data. The declining dotted line reflects mostly the shift of intercity travel from cars to airplanes. Significant events, mentioned in the text, which influence the trends, are pointed out. WWII affected both car deaths and gun deaths similarly in an obvious way.

**2.3 The recent increase of the number of deaths**

From the early 2010s onward the number of deaths from guns but also from car accidents began increasing again, hand in hand, see Figure 2. Among the possible explanations for the causes of the rising fatalities could be: increased gun ownership, decreased gun control measures (in some states,) the rising drug addiction rates, and the rise of social media and the decline in our sense of community. And yet, there is no reason to believe that the recent increase in fatalities will develop into major long-term trend extending far above the long-established equilibrium level of around 10 per 100,000 annually, which has been auto-regulated over extended periods of time. It is more reasonable to consider this deviation as the beginning of a short-lived fluctuation that will be eventually reabsorbed like past ones. Public outcry may trigger new legislative actions, or a shift of societal behaviors in that direction. For example, President Biden promised that ending America's gun violence epidemic was "within our grasp" and on April 8th 2021 he laid out his first steps to bring the problem of "ghost" guns under control. On June 25, 2022 he signed into law the Bipartisan Safer Communities Act (BSCA). These, plus other possible similar legislative actions, are posed to reverse the recently rising trend.

The number of fatalities from guns and also from car accidents are likely to soon stop growing and begin tracing bell-shaped trajectories, like the smooth gray lines shown in Figure 2. These curves depict the life cycles (the rates of growth, the derivatives) of logistic curves fitted on the numbers of "excess" fatalities since 2011. The data are fitted on the expression of the time derivative of Equation 1, namely:

$$f'(t) = \frac{\alpha M}{(1+e^{-\alpha(t-t_0)})(1+e^{\alpha(t-t_0)})} \qquad (2)$$

Table 1 gives the parameters as determined by the fits. In doing so we are treating such deviations as local, small-scale natural-growth processes, limited in time, rather than a new phenomenon with the capacity to grow and establish an eventually higher equilibrium level for the number of annual fatalities. The decision to consider such micro niches can be justified in view of society's tendency to self-regulate safety as manifested in the past.

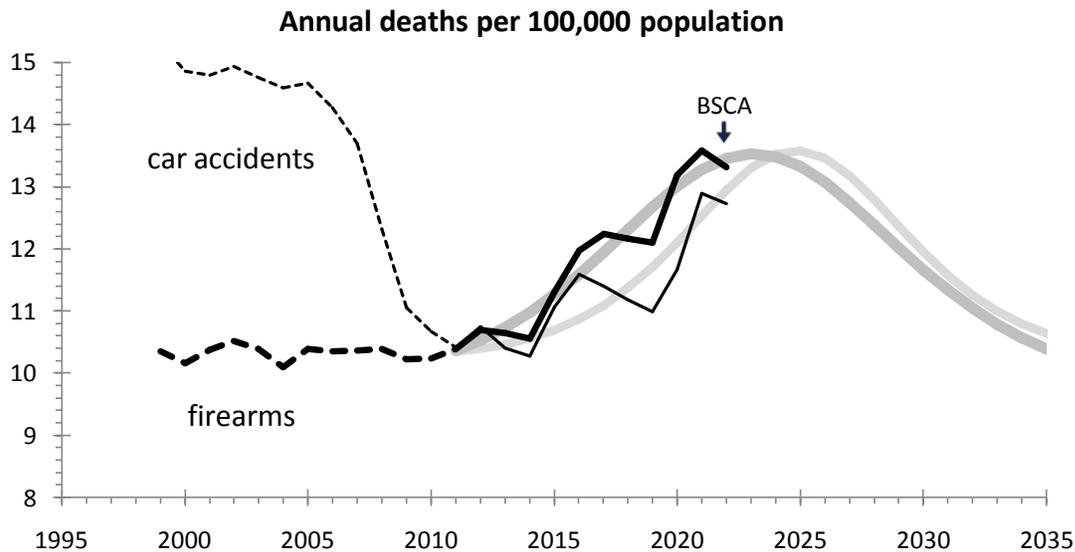

Data sources: NHTSA[8]; Gun Violence Archive[9]

Figure 2. Annual rates of deaths from car accidents and from the unlawful use of guns during recent decades. The smooth gray lines are the life cycles (rates of growth) the derivatives of logistic functions fitted to the solid black lines. The enactment of the Bipartisan Safer Communities Act (BSCA) mentioned in the text is pointed out.

### 3. Mass shootings

However, mass-shooting deaths are a different matter.[1] Mass shootings produce abundant headlines and vocal reaction from the public because of the emotions they trigger. But they are responsible for only 0.1 percent of all deaths from the unlawful use of guns. This is one of the reasons that legislative measures taken in the past had little impact on this number. Also, the evolution of the number of fatalities from mass shootings since 1960 is very different from the evolution of the overall firearm fatalities discussed earlier. Mass-shooting deaths are growing exponentially rather than along a flat horizontal pattern like that of gun deaths.

Society's corrective actions may often seem to be triggered by the publicity given to mass shootings, but the fundamental reason for society's response is an upward excursion of the *overall* number of gun deaths, which is generally eliminated afterward. Gun-control measures have had little effect on the evolution of mass shootings. Loud public outcry may render people more careful with their guns but it will not deter potential mass shooters from putting their sinister contemplations into actions. On the contrary, extensive publicity will spread the word further and will incite other such perpetrators elsewhere in the country.

Mass shootings seem to behave as a "species" different from other firearm fatalities and as such it should be described with a growth curve of their own. The data shown in Figure 3 – black line – are 5-year averages because the year-to-year fluctuations are large. The logistic curve fitted on these data – smooth gray line – is still in its very early stages, which makes the pattern indistinguishable from a simple exponential pattern. In such cases a final-ceiling for the logistic curve cannot be reliably estimated. One thing is certain; the final death rate from mass shootings will be much higher than today's toll. But by how much and by when depends on what will happen between now and then.

---

[1] The definition of a mass public shooting used here is the same as that at The Violence Project and is given by the Congressional Research Service as follows:
"a multiple homicide incident in which four or more victims are murdered with firearms—not including the offender(s)—within one event, and at least some of the murders occurred in a public location or locations in close geographical proximity (e.g., a workplace, school, restaurant, or other public settings), and the murders are not attributable to any other underlying criminal activity or commonplace circumstance (armed robbery, criminal competition, insurance fraud, argument, or romantic triangle)."

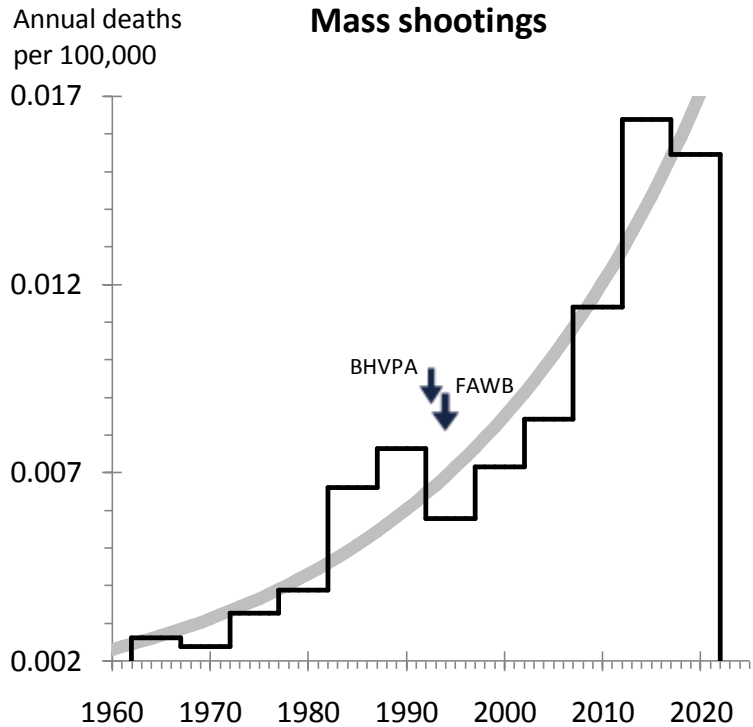

Data source: The Violence Project[10]

Figure 3. Five-year averages of the annual rates of deaths during mass shootings. The smooth gray line is a logistic fit to the data, but so far it is indistinguishable from an exponential pattern. The Brady Handgun Violence Prevention Act of 1993 (BHVPA) and the Federal Assault Weapons Ban of 1994 (FAWB) had minimal effect on the evolution of this trend.

**Table I**
**Fit parameters for all the logistic fits**

|  | cars | guns | Mass shootings | car micro niche | gun micro niche |
|---|---|---|---|---|---|
| $M$ | 22.5 | 10.6 | 0.24* | 41.0 | 51.8 |
| $\alpha$ | 0.29 | 0.2 | 0.038* | 0.33 | 0.25 |
| $t_0$ | 1919.3 | 1913.5 | 2087.6* | 2024.8 | 2023.1 |
| Completion level of logistic function | 93% | 96% | 8%* | 28% | 35% |
| Correlation at completion level | 99% | 99% | 95% | 97% | 89% |

* These numbers are very uncertain. For example, the error on $M$=0.24 could be significantly greater than ±500% with confidence level 90%, and the error on $t_0$=2087 significantly greater than ±2%, according to the tables provided by Debecker and Modis.[11]

## 4. Discussion

Table 1 gives all the parameters as they have been determined by the various fitting procedures. A "completion level" is calculated for each logistic function as the latest point covered by data on the sigmoid path before the final equilibrium level is reached. Correlations up to the completion level are given to provide some measure of the goodness of the fit. The uncertainties of the parameters determined for the mass-shootings fit are very large because the penetration level is only 8%. At present this translates to $t \ll t_0$ in Equation 1, which makes the logistic function effectively indistinguishable from a simple exponential function.

In fact it is not surprising that mass shootings have been growing exponentially. Mass shootings have often been likened to an epidemic, and the most outstanding characteristic of an epidemic is that it begins with exponential growth. Each mass shooting inspires potential mass shooters, triggering more such actions sooner or later. In other words, mass shootings have the ability to multiply like a species: the rate of appearance of new ones is proportional to how many there have been already, which spells out exponential growth. Professors James Densley (sociologist) and Jillian Petereson (psychologist) argue "Each new [mass] shooting normalizes the process and encourages new participants to join in."[12] In the absence of an effective restraining mechanism this number will continue growing that way.

The smooth gray exponential curve in Figure 3 is the early beginning of a natural-growth logistic curve. Natural-growth curves proceed to completion as long as the conditions remain unchanged. What is meant here by "the conditions remain unchanged" is that we will be seeing in the future the *same kind* of restraining efforts as we saw in the past. No deviation from the projected curve can be expected if only small incremental legislative actions take place on state or federal level, like the ones in the 1990s, or the ones enacted by Biden. Such type of actions has already influenced the evolution of the data and consequently has been accounted for in the calculation of the growth curve. To observe a deviation from this trajectory we must witness something *unlike* anything seen before, for example, some change in the American constitution, or alternatively, Weapons Bans more severe than the one which expired in 2004. Short of actions of this magnitude we must expect "business as usual" namely the continuation of the curve as it has been estimated. This would eventually render mass shootings a menace to society comparable to suicide and other homicide causing the overall number of fatalities to change trend from a flat homeostatic level into an exponentially growing one.

When this happens, late in this century, the overall number of gun deaths will rises well above 10 per 100,000 annually, and society will definitely react again – if it hasn't already done so before then – and this time with unprecedented action that will alter the exponential trend into a logistic pattern. However, waiting until we reach this point may come at a great cost in human lives.

It is worth noting that an annual death rate of 10 per 100,000 inhabitants is characteristic of American society. In contrast, Europe's equivalent number of deaths from car accidents is less than 5, and that from gun deaths less than 1.

The author declares there is no conflict of interest.

**References**

[1] Herbert Spencer, The Principles of Sociology, Greenwood, 1975. First published in 1877.

**Theodore Modis** is a physicist, strategist, futurist, and international consultant. He is author/co-author to over one hundred articles in scientific and business journals and ten books. He has on occasion taught at Columbia University, the University of Geneva, at business schools INSEAD and IMD, and at the leadership school DUXX, in Monterrey, Mexico. He is the founder of Growth Dynamics, an organization specializing in strategic forecasting and management consulting: http://www.growth-dynamics.com